\documentstyle[twoside,psfig]{article}

\catcode`\@=11
\long\def\@makefntext#1{
\protect\noindent \hbox to 3.2pt {\hskip-.9pt
$^{{\eightrm\@thefnmark}}$\hfil}#1\hfill}               

\def\@makefnmark{\hbox to 0pt{$^{\@thefnmark}$\hss}}    

\def\ps@myheadings{\let\@mkboth\@gobbletwo
\def\@oddhead{\hbox{}
\rightmark\hfil\eightrm\thepage}
\def\@oddfoot{}\def\@evenhead{\eightrm\thepage\hfil
\leftmark\hbox{}}\def\@evenfoot{}
\def\sectionmark##1{}\def\subsectionmark##1{}}

\oddsidemargin=\evensidemargin
\newcommand{\beq}{\begin{equation}}
\newcommand{\eeq}{\end{equation}}

\newcounter{sectionc}\newcounter{subsectionc}\newcounter{subsubsectionc}
\renewcommand{\section}[1] {\vspace{12pt}\addtocounter{sectionc}{1}
\setcounter{subsectionc}{0}\setcounter{subsubsectionc}{0}\noindent
        {\tenbf\thesectionc. #1}\par\vspace{5pt}}
\renewcommand{\subsection}[1] {\vspace{12pt}\addtocounter{subsectionc}{1}
        \setcounter{subsubsectionc}{0}\noindent
        {\bf\thesectionc.\thesubsectionc. {\kern1pt \bfit #1}}\par\vspace{5pt}}
\renewcommand{\subsubsection}[1] {\vspace{12pt}\addtocounter{subsubsectionc}{1}
        \noindent{\tenrm\thesectionc.\thesubsectionc.\thesubsubsectionc.
        {\kern1pt \tenit #1}}\par\vspace{5pt}}

\newcounter{appendixc}
\newcounter{subappendixc}[appendixc]
\newcounter{subsubappendixc}[subappendixc]
\renewcommand{\thesubappendixc}{\Alph{appendixc}.\arabic{subappendixc}}
\renewcommand{\thesubsubappendixc}
        {\Alph{appendixc}.\arabic{subappendixc}.\arabic{subsubappendixc}}

\renewcommand{\appendix}[1] {\vspace{12pt}
        \refstepcounter{appendixc}
        \setcounter{figure}{0}
        \setcounter{table}{0}
        \setcounter{lemma}{0}
        \setcounter{theorem}{0}
        \setcounter{corollary}{0}
        \setcounter{definition}{0}
        \setcounter{equation}{0}
        \renewcommand{\thefigure}{\Alph{appendixc}.\arabic{figure}}
        \renewcommand{\thetable}{\Alph{appendixc}.\arabic{table}}
        \renewcommand{\theappendixc}{\Alph{appendixc}}
        \renewcommand{\thelemma}{\Alph{appendixc}.\arabic{lemma}}
        \renewcommand{\thetheorem}{\Alph{appendixc}.\arabic{theorem}}
        \renewcommand{\thedefinition}{\Alph{appendixc}.\arabic{definition}}
        \renewcommand{\thecorollary}{\Alph{appendixc}.\arabic{corollary}}
        \renewcommand{\theequation}{\Alph{appendixc}.\arabic{equation}}
        \noindent{\tenbf Appendix \theappendixc #1}\par\vspace{5pt}}
\newcommand{\subappendix}[1] {\vspace{12pt}
        \refstepcounter{subappendixc}
        \noindent{\bf Appendix \thesubappendixc. {\kern1pt \bfit #1}}
        \par\vspace{5pt}}
\newcommand{\subsubappendix}[1] {\vspace{12pt}
        \refstepcounter{subsubappendixc}
        \noindent{\rm Appendix \thesubsubappendixc. {\kern1pt \tenit #1}}
        \par\vspace{5pt}}

\newcommand{\textlineskip}{\baselineskip=13pt}
\newcommand{\smalllineskip}{\baselineskip=10pt}

\def\eightcirc{
\begin{picture}(0,0)
\put(4.4,1.8){\circle{6.5}}
\end{picture}}
\def\eightcopyright{\eightcirc\kern2.7pt\hbox{\eightrm c}}

\def\keywords#1{{
        \centering{\begin{minipage}{4.5in}\footnotesize\baselineskip=10pt
        {\footnotesize\it Keywords}\/: #1
         \end{minipage}}\par}}

\renewenvironment{thebibliography}[1]
        {\frenchspacing
         \ninerm\baselineskip=11pt
         \begin{list}{\arabic{enumi}.}
        {\usecounter{enumi}\setlength{\parsep}{0pt}     
         \setlength{\leftmargin 12.7pt}{\rightmargin 0pt} 
         \setlength{\itemsep}{0pt} \settowidth
        {\labelwidth}{#1.}\sloppy}}{\end{list}}

\newcounter{itemlistc}
\newcounter{romanlistc}
\newcounter{alphlistc}
\newcounter{arabiclistc}

\newcommand{\fcaption}[1]{
        \refstepcounter{figure}
        \setbox\@tempboxa = \hbox{\footnotesize Fig.~\thefigure. #1}
        \ifdim \wd\@tempboxa > 5in
           {\begin{center}
        \parbox{5in}{\footnotesize\smalllineskip Fig.~\thefigure. #1}
            \end{center}}
        \else
             {\begin{center}
             {\footnotesize Fig.~\thefigure. #1}
              \end{center}}
        \fi}

\newcommand{\tcaption}[1]{
        \refstepcounter{table}
        \setbox\@tempboxa = \hbox{\footnotesize Table~\thetable. #1}
        \ifdim \wd\@tempboxa > 5in
           {\begin{center}
        \parbox{5in}{\footnotesize\smalllineskip Table~\thetable. #1}
            \end{center}}
        \else
             {\begin{center}
             {\footnotesize Table~\thetable. #1}
              \end{center}}
        \fi}

\def\@citex[#1]#2{\if@filesw\immediate\write\@auxout
        {\string\citation{#2}}\fi
\def\@citea{}\@cite{\@for\@citeb:=#2\do
        {\@citea\def\@citea{,}\@ifundefined
        {b@\@citeb}{{\bf ?}\@warning
        {Citation `\@citeb' on page \thepage \space undefined}}
        {\csname b@\@citeb\endcsname}}}{#1}}

\newif\if@cghi
\def\cite{\@cghitrue\@ifnextchar [{\@tempswatrue
        \@citex}{\@tempswafalse\@citex[]}}
\def\citelow{\@cghifalse\@ifnextchar [{\@tempswatrue
        \@citex}{\@tempswafalse\@citex[]}}
\def\@cite#1#2{{$\null^{#1}$\if@tempswa\typeout
        {IJCGA warning: optional citation argument 
        ignored: `#2'} \fi}}

\def\pmb#1{\setbox0=\hbox{#1}
        \kern-.025em\copy0\kern-\wd0
        \kern.05em\copy0\kern-\wd0
        \kern-.025em\raise.0433em\box0}

\def\fnt#1#2{\footnotetext{\kern-.3em
        {$^{\mbox{\scriptsize #1}}$}{#2}}}

\def\ps@myheadings{%
    \let\@oddfoot\@empty\let\@evenfoot\@empty
    \def\@evenhead{\slshape\leftmark\hfil}
    \def\@oddhead{\hfil{\slshape\rightmark}}
    \let\@mkboth\@gobbletwo
    \let\sectionmark\@gobble
    \let\subsectionmark\@gobble
    }
\font\tenrm=cmr10
\font\tenit=cmti10 
\font\tenbf=cmbx10
\font\bfit=cmbxti10 at 10pt
\font\ninerm=cmr9

\font\eightrm=cmr8

\def\qed{\hbox{${\vcenter{\vbox{                        
   \hrule height 0.4pt\hbox{\vrule width 0.4pt height 6pt
   \kern5pt\vrule width 0.4pt}\hrule height 0.4pt}}}$}}

\begin{document}
\setlength{\textheight}{7.7truein}  

\thispagestyle{empty}

\normalsize\textlineskip

\setcounter{page}{1}

\vspace*{0.88truein}

\centerline{\bf Energy Dependence of the $\Lambda/\Sigma ^0$ Production in Proton-Proton Collisions}
\centerline{\bf Near the Hyperon-$K^+$ Thresholds\footnote{Supported in part by Forschungszentrum FZ J\"ulich (COSY)}}\baselineskip=13pt
\baselineskip=13pt
\vspace*{0.4truein}
\centerline{\footnotesize M. Dillig\footnote{Email: mdillig@theorie3.physik.uni-erlangen.de}  and M. Schott}
\baselineskip=12pt
\centerline{\footnotesize\it Institute for Theoretical Physics III
\footnote{preprint FAU-TP3-06/Nr. 03}}
\centerline{\footnotesize\it University of
Erlangen-N\"urnberg}  
\centerline{\footnotesize\it Staudtstr. 7, Erlangen, D-91058, Germany}
\vspace*{12pt}
\vspace*{0.23truein} 
\begin{quotation}
The energy dependence of the reactions $pp \rightarrow p \Lambda K^+$ and $pp \rightarrow p \Sigma ^0 K^+$ and the ratio $R _{\Lambda / \Sigma^0}$ is studied in a constituent quark-gluon model, including the excitation of the baryon resonances $N^*(1650)$, $N^*(1710)$ and $N^*(1720)$ near the $\Lambda / \Sigma ^0$ thresholds. Representing the baryons as quark-diquark objects, the energy dependent ratio $R_{\Lambda/\Sigma^0}$, which is qualitatively reproduced up to excess energies of 60 MeV above threshold, provides detailed information on the momentum spectrum of axial diquarks in the proton and the $\Sigma^0$.
\end{quotation}
\vspace*{12pt}
\vspace*{2pt}
\normalsize                     
\vspace*{-0.5pt}
\noindent

\bigskip
\vspace*{12pt}

A large fraction of physics at COSY is the investigation of exclusion heavy meson production in $pp \rightarrow p B \lambda$ near their corresponding thresholds ($\lambda$ refers to different mesons with masses $\leq 1GeV$ [1]). In particular associated strangeness into $\Lambda, \Sigma$ final states [2] has been measured at COSY-11 ([3]-[5]) and COSY-TOF [6]. The most striking feature of the data is the ratio

\begin{equation}R _{\Lambda / \Sigma ^0} (Q) = \frac{\sigma(pp \rightarrow p \Lambda K^+)}{\sigma(pp \rightarrow p \Sigma ^0 K^+)} \end{equation}

(as a function of the excess energy $Q = \sqrt{s} - (M_P + M_Y + m_{K^+})$ with $Y = \Lambda, \Sigma ^0$ and $\sqrt{s}$ as the center-of-mass energy) as shown in Fig. 1: close to threshold with $Q \leq 10MeV$ the $\Sigma ^0$ is suppressed relative to the $\Lambda$ by a factor 30 and at $Q\leq 60MeV$ still by one order of magnitude. Opposite, for large $Q>300MeV$ the $\Lambda / \Sigma^0$ ratio approaches $\sim 2.5$ [7], as expected from isospin relations.\\

So far, the ratio $R_{\Lambda/\Sigma^0}(Q)$ has been investigated in two classes of meson-exchange models. In particular the models of the J\"ulich group involve $\pi$ and $K$-exchange, where their relative phase interferes constructively or destructively in the coherent sum ([8]-[10]); or $\pi$ and K exchange are summed incoherently. In the second class of models the reaction mechanisms are dominated by heavy meson exchange together with the excitation of $N^*$ resonances around the $\Lambda, \Sigma ^0$ thresholds ([11]-[15]). The different results, summarized in ([1][2]), point out that all models (with constructive $\pi K$ interference in the J\"ulich model) qualitative reproduce the decrease of $R_{\Lambda /  \Sigma^0}(Q)$ with increasing excess energies, without allowing, however, quantitative conclusions on the dominant production mechanism.\\

In this note we approach $\Lambda$ and $\Sigma ^0$ hyperon production differently: in view of the typical momentum transfer of $\Delta q \sim 900$ MeV/c even at threshold, we model the strong overlapping hadrons in terms of the constituent quark and gluon degrees of freedom.\\


Thereby we expand the production operator in terms of multiple gluon exchanges up $\alpha _s^3$, with $\alpha _s$ being the strong coupling constant at the squared $qqg$ vertex (contributions from the 3-gluon interaction ggg are so far not included). With the standard one-gluon exchange (QGE) qq interaction [16]

\begin{equation}V_{ij}^{qq} (q) = 4 \pi \alpha _s \frac{\vec \lambda ^a _j  \vec \lambda ^a _i}{4} \frac{\gamma _\mu (i) \gamma^{\mu} (i)}{q^2 - m_q^2 - i \epsilon} \end{equation}

 the qq interaction, upon reduction of the corresponding Dirac spinors, involves central, spin-spin, spin-orbit and tensor terms [16][17], while the excitation of the $K^+$ meson in $q \rightarrow q (q \bar q)$ leads to([17][18])

\begin{equation}V _{q \rightarrow q (q \bar q)} (\vec q) = 4 \pi \alpha_s \frac{\vec \lambda _i^a \vec\lambda _j ^a}{4} \frac{1}{m _q m _g^2} \left( (\vec \sigma _i \times \vec \sigma _j) \vec q + 2 \vec \sigma _j \vec p _i \right) \end{equation}

As our investigation focuses on a very restricted range of momentum transfers we drop the smooth momentum dependence of the various effective parameters with $\alpha _s \sim 2$, $m _q \sim 330$ MeV and $m _g \sim 800$ MeV, as obtained from hadron spectroscopy in improved non relativistic quark model (keeping the leading nonstatic corrections at the vertices). Among the various contributions in powers of $\alpha _s$ [19], the leading term is the exchange of a color singlet $gg$-pair, followed by the QGE induced $q\bar q$ excitation Ã¢ÂÂ as shown in Fig. 2 (together with the corresponding diagram in meson exchange models); it quite naturally leads to the excitations of intermediate baryon resonances. In this calculation we include the $N^*(1650)$, $N^*(1710)$ and $N^*(1720)$ around the $\Lambda, \Sigma ^0$ thresholds [20].\\


The crucial input for the transition amplitude is the simplification of the $6q$ and $6q-(q\bar q)$ many body problem for the pp and the $ p Y K^+$ systems, respectively: we represent all baryons as a $q-(qq)$ diquark systems. Classifying the proton from its spin-isospin (flavor) structure, it is a superposition of a scalar $S=T=0$ and an axial $S=T=1$ diquark

\begin{eqnarray}
|p> \sim C_S \left[ \frac{1}{2} [\frac{1}{2} \frac{1}{2}]^{S=T=0} \right]^{ \frac{1}{2} \mu, \frac{1}{2} \frac{1}{2}} + C_A \left[ \frac{1}{2} [\frac{1}{2} \frac{1}{2}]^{S=T=1} \right]^{ \frac{1}{2} \mu, \frac{1}{2} \frac{1}{2}} 
\end{eqnarray}

(for all quarks being in relative $0s$ states) with $C_S = C_A = \frac{1}{\sqrt{2}}$ without dynamical correlations in the diquark systems. Now, different approaches to the quark structure of the proton, such as on the lattice ([21][22]), the Dyson-Schwinger and Bethe-Salpeter framework ([23]-[26]) and in covariant ([27][28]) and non relativistic quark models ([29][30]) all confirm much stronger correlations for the scalar versus the axial diquark: most observables of the proton are semi-quantitatively described in keeping only scalar diquarks. Following ref. [30], typically $(C_S^2 \sim 0.95) \sim (10-20) (C_A^2 \sim 0.05)$ with $m_S \sim 600MeV < 2 m_{quark} \ll m_A \sim 800MeV$.\\

$\Lambda / \Sigma^0$ strangeness production provides an excellent opportunity towards a detailed diquark structure of the proton: the scalar nature of the $\Lambda$

\begin{equation} |\Lambda> = \Phi _S (r) \left[ s [u d]^{S=T=0}\right]^{\frac{1}{2} \mu , 00} \end{equation}

and the axial structure of the $\Sigma ^0$

\begin{equation} |\Sigma ^0> = \Phi _A (r) \left[ s [u d]^{S=T=1}\right]^{\frac{1}{2} \mu , 10} \end{equation}

select (non-flavor changing) transitions from the proton. As we test the corresponding wave functions at very large momentum transfers, we include a d-state admixture in the axial diquark, as expected from the tensor force in the OGE and from the Bethe-Salpeter equation for the $qq$ system. This yields for the proton

\begin{eqnarray} 
|p> =& &\left( C_S \Phi _S (r, \phi) \left[\frac{1}{2}[\frac{1}{2} \frac{1}{2}]^{00} \right] \right.
\\
& & + \sum _{L=0,2; S=0,1} C_{AL} \Phi _{AL} (r, \phi) 
\left. \left[ Y _L (\hat {\vec r})  [\frac{1}{2} [\frac{1}{2} \frac{1}{2}]^{11}]^S \right]^{ \frac{1}{2} \mu, \frac{1}{2} \frac{1}{2}} \right) \frac{\delta _{ij}}{\sqrt{3}}
\end{eqnarray} 

where we estimate perturbatively $C^2_{A2} \sim 0.1 C _{A0}$ ($\delta _{ij}/\sqrt{3}$ represent the q-diquark color content).\\

Typical results for the momentum spectra $C_S^2|\Phi _S (q)|^2$, $C_A^2 |\Phi _A (q)|^2$ with $C_S^2 = 0.95$, $C_{A0}^2 = 0.05$ and $C^2_{A2} / C^2_{A0} = 0.1$ are shown in Fig. 3 for Gaussian size parameters $a_S$ = 0.5 fm and $a_A$ = 0.55 fm. Clearly, at large q the $L=2$ component in the axial diquark becomes increasingly important, though, in view of significant uncertainties of model calculation of the diquarks, Fig. 3 presents only a qualitative view of the momentum spectra of diquarks. \\


With the ingredients above the observables, the $pp \rightarrow p \Lambda (\Sigma ^0) K^+$ cross sections and their ratio, are readily calculated from

\begin{equation} \sigma (Q) = \frac{1}{(2 \pi)^5 v_{pp}} \frac{1}{4} \int{\sum_{spin} |<f|V_{qq \rightarrow qq (q \bar q)}|i>|^2 dV_{ps}}\end{equation}

with the initial state

\begin{equation} |i> = \sqrt{\frac{\sigma _{pp}^{el}}{\sigma _{pp}^{tot}}} \Phi _{pp} ( r,  \rho,  r' , \rho') \, e^{i \vec p \vec R}\end{equation}

(for Jacobi coordinates $r, \rho,  r',  \rho', r'', \vec R$, with $\vec R$ being the relative distance between the two baryons),
whereby the ratio of the elastic to the total $pp$ cross section [20] estimates the reduction of the cross section from the coupling of the pp system to inelastic channels in initial state interactions and the final state

\begin{equation}  |f> = \Phi _{pY} (r, \rho, r', \rho') \Phi _\omega (r'') \chi _{pY} (\vec R) \end{equation}

where the distorted wave $\Phi_{PY}$ incorporates the pY final state interactions, parametrized in terms of the scattering length $a_{pY}$ and the effective range $r_{pY}$, respectively (the very weak $K^+$ baryon final state interaction is not included ([31]-[33])).\\ 

For the practical calculations, all multiple integrals are evaluated in a Gaussian expansion of both the bound state, the pY distorted wave and the transition operator; only the final one dimensional integration over phase space $dV_{ps}$ is performed numerically.\\

\vspace*{13pt}
%

Fig. 4 and 5 show a comparison with experimental data. In Fig. 4 the energy dependence of the total $pp \rightarrow p \Lambda K^+$ and $pp \rightarrow p \Sigma ^0 K^+$ cross sections is reproduced qualitatively. Though the results shows a significant sensitivity on the diquark parameters, firm conclusions are prevented by the dominance of phase space and pY final state interaction; furthermore, the very qualitative account of pp initial state interactions, which reduce the total cross sections by typically a factor 0.4, prevents an absolute normalization of the predictions.\\

Fig. 5 presents the energy dependence of $R _{\Lambda / \Sigma^0} (Q)$: as the ratio is less sensitive to phase space and initial and final state interactions, is should reveal more stringent information on diquarks. Keeping all additional parameters fixed, the ratio is sensitivity influenced by the scalar versus axial content of the proton and by their size parameters. Without searching for a best fit of the data, they can be reproduced nearly quantitatively with $c_s^2 / c_{A0}^2 / c_{A2}^2 = 95 / 5 / 0.5 $ and $a_S$ = 0.9 $a_A$ = 0.6 fm. Even being very cautious in our conclusions, the result confirms qualitatively a very large scalar diquark content with a probability of around $95\%$ versus a strength of typically $5\%$ of the axial diquark in the proton [30]; it excludes the result from ref. [34] with $\sim 20\%$ of axial diquark.\\

It is clear that for further insight, additional experimental information is required. Particularly interesting would be detailed information on angular distributions for both $\Lambda$ and $\Sigma ^0$ the the final states, the excitation of $p \Sigma K$ final states into different isospin channels and an extension of $R _{\Lambda / \Sigma ^0} (Q)$ to $Q > 60MeV$ for a transition to excess energies up to $Q \sim 300$ MeV. For theoretical progress, a detailed baryon spectroscopy in a (semi) relativistic quark-diquark model has to be performed. Beyond that, most promising will be a consistent description of $\Lambda (\Sigma) K$ excitation in the $pp$ and in $\gamma p$ on the nucleon ([35][36]), as in photo-induced hyperon production both initial and final state interactions are practically absent, so that calculations with absolute normalization will be feasible.\\

Summarizing, already this first analysis in a quark-gluon model reveals new information on proton-induced $\Lambda, \Sigma$ production: their comparative study provides a unique filter for details of the proton structure and on the clustering of qq systems as scalar and axial diquarks.

\newpage



%

\newpage


\begin{figure}[htbp] 
\vspace*{13pt}
\framebox{\centerline{\psfig{width=10.0cm,file=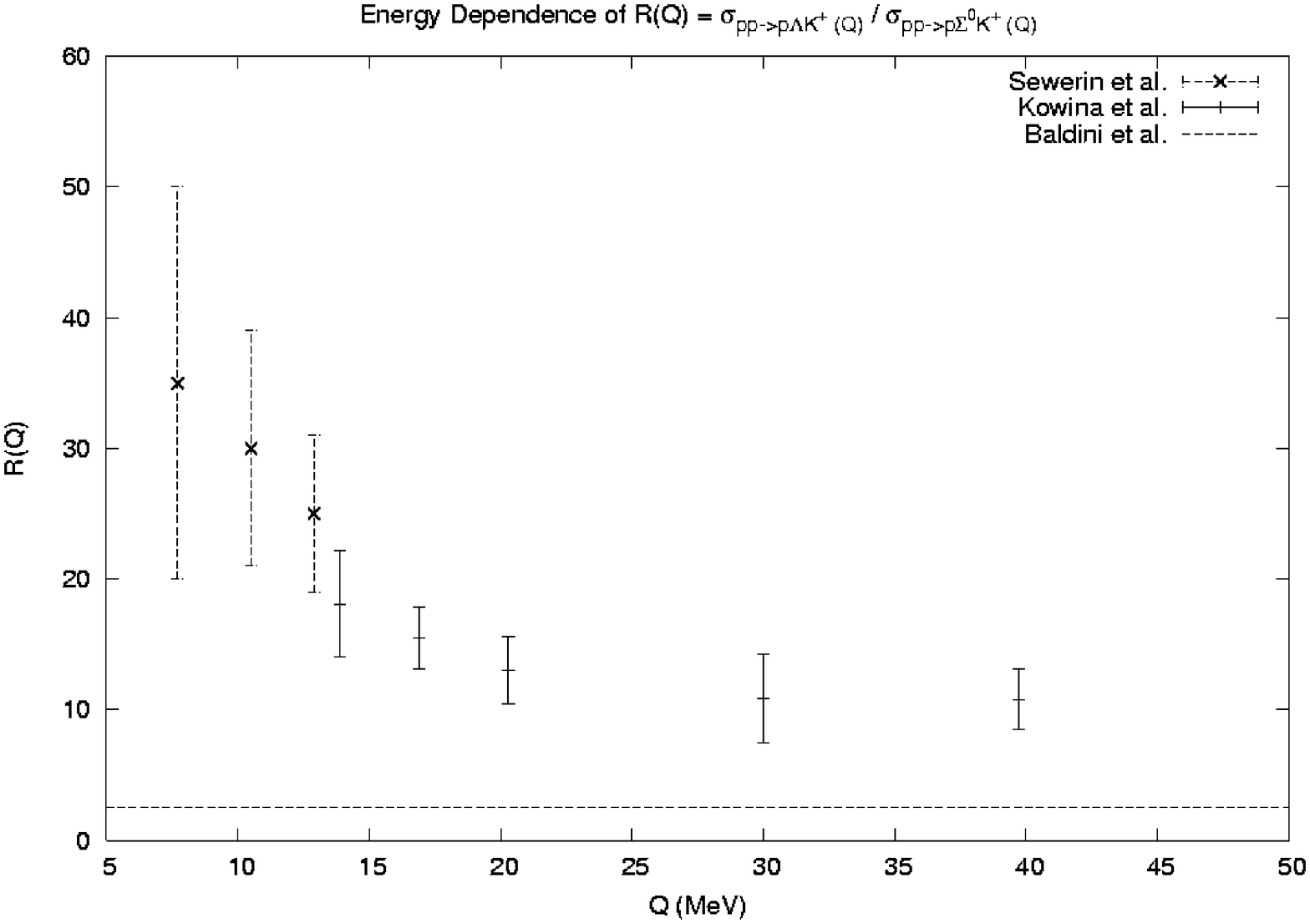}}} 
\vspace*{13pt}
\vspace{-0.5cm}
\fcaption{Ratio $R _{\Lambda / \Sigma ^0}$ of the cross sections $pp\rightarrow p K^+ \Lambda, p K^+ \Sigma ^0$ as a function of the excess energy $Q=\sqrt{s}-(M_p + M_\Upsilon) + m _{K^+})$, with $\sqrt{s}$ as the center of mass energy. The data at $Q<60MeV$ are from refs. ([3]-[5]). For $Q>300MeV$ the data point from [7] is indicated by the dashed line}
\end{figure}

\begin{figure}[htbp] 
\vspace*{13pt}
\framebox{\centerline{\psfig{width=10.0cm,file=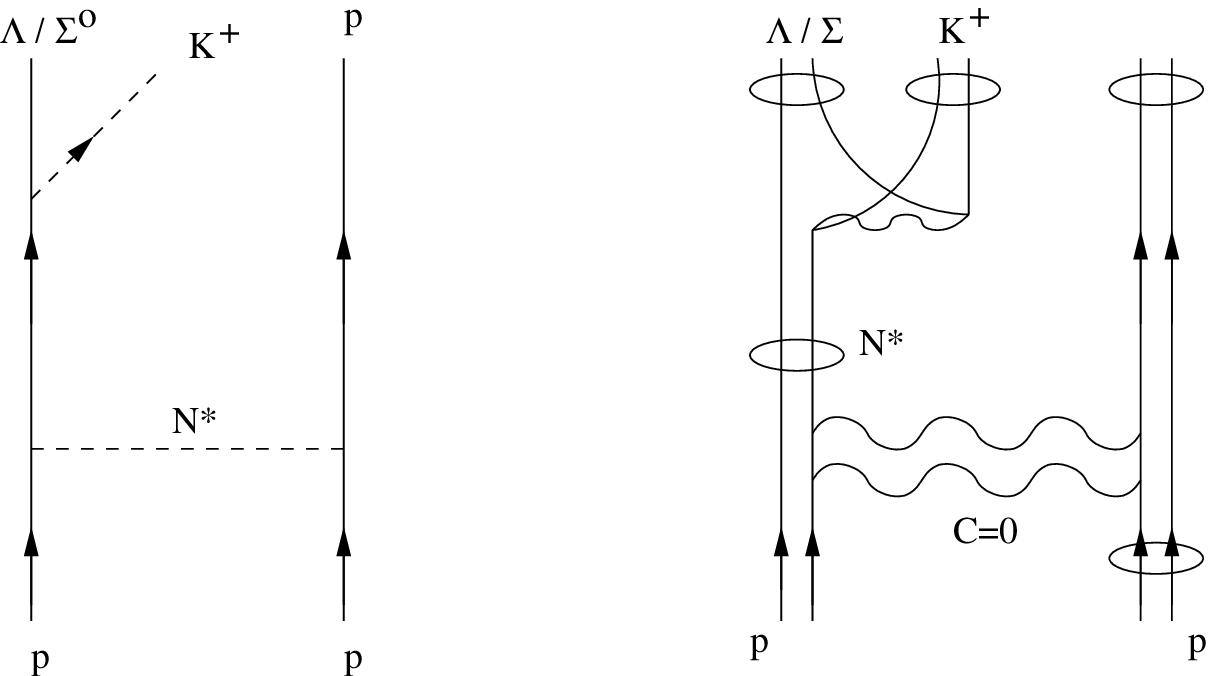}}} 
\vspace*{13pt}
\vspace{-0.5cm}
\fcaption{Excitation of baryon resonances in meson exchange (a) and quark-gluon models (b)}
\end{figure}

\begin{figure}[htbp] 
\vspace*{13pt}
\framebox{\centerline{\psfig{width=10.0cm,file=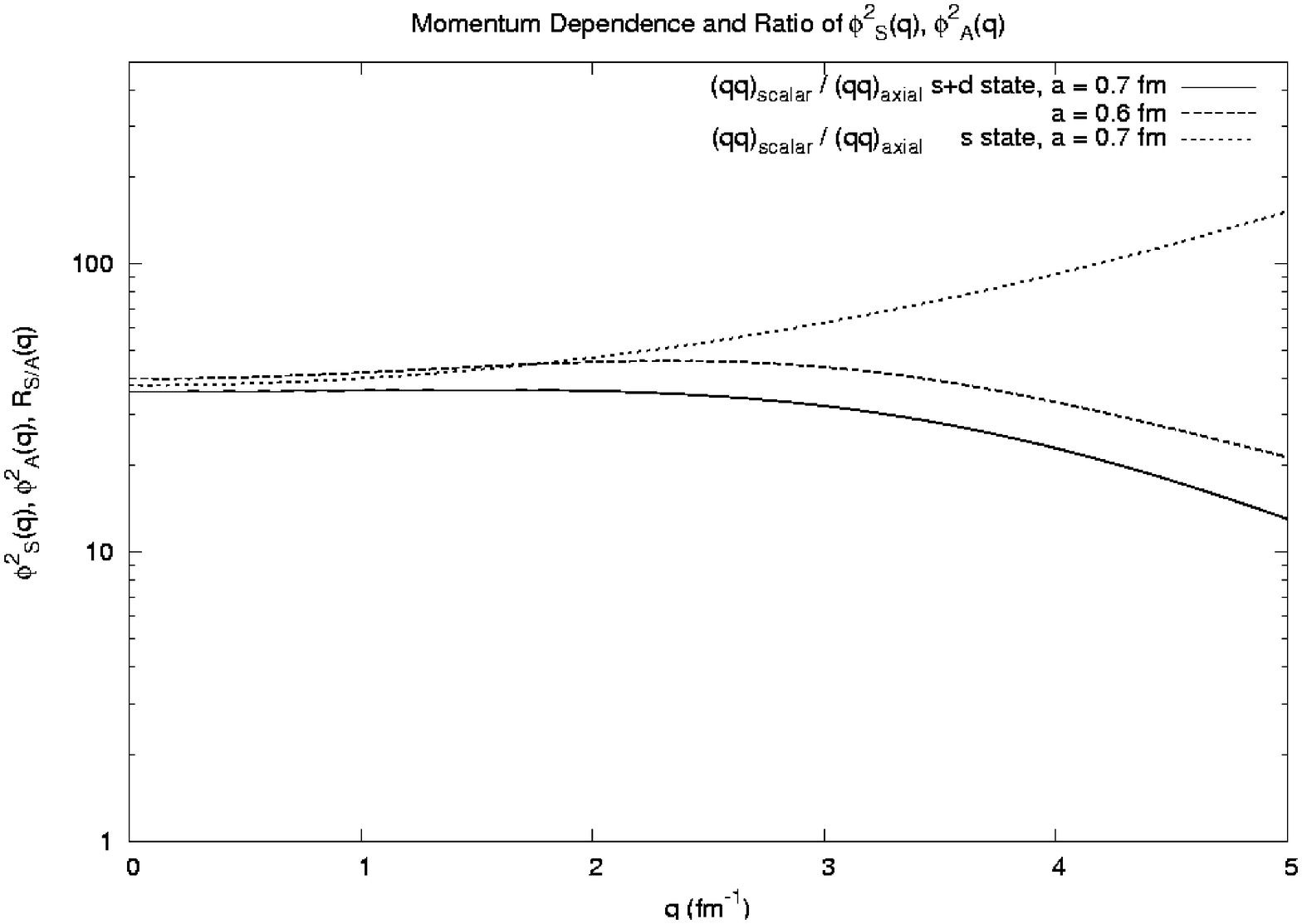}}} 
\vspace*{13pt}
\vspace{-0.5cm}
\fcaption{Momentum spectrum of the ratio $|\Phi _s (q)|^2$ versus $|\Phi _A (q)|^2$ of scalar and axial diquarks (with a d-state with a probability $P_D$ = 0.1 and two size parameters) and for pure s-states}
\end{figure}

\begin{figure}[htbp] 
\vspace*{13pt}
\framebox{\centerline{\psfig{width=10.0cm,file=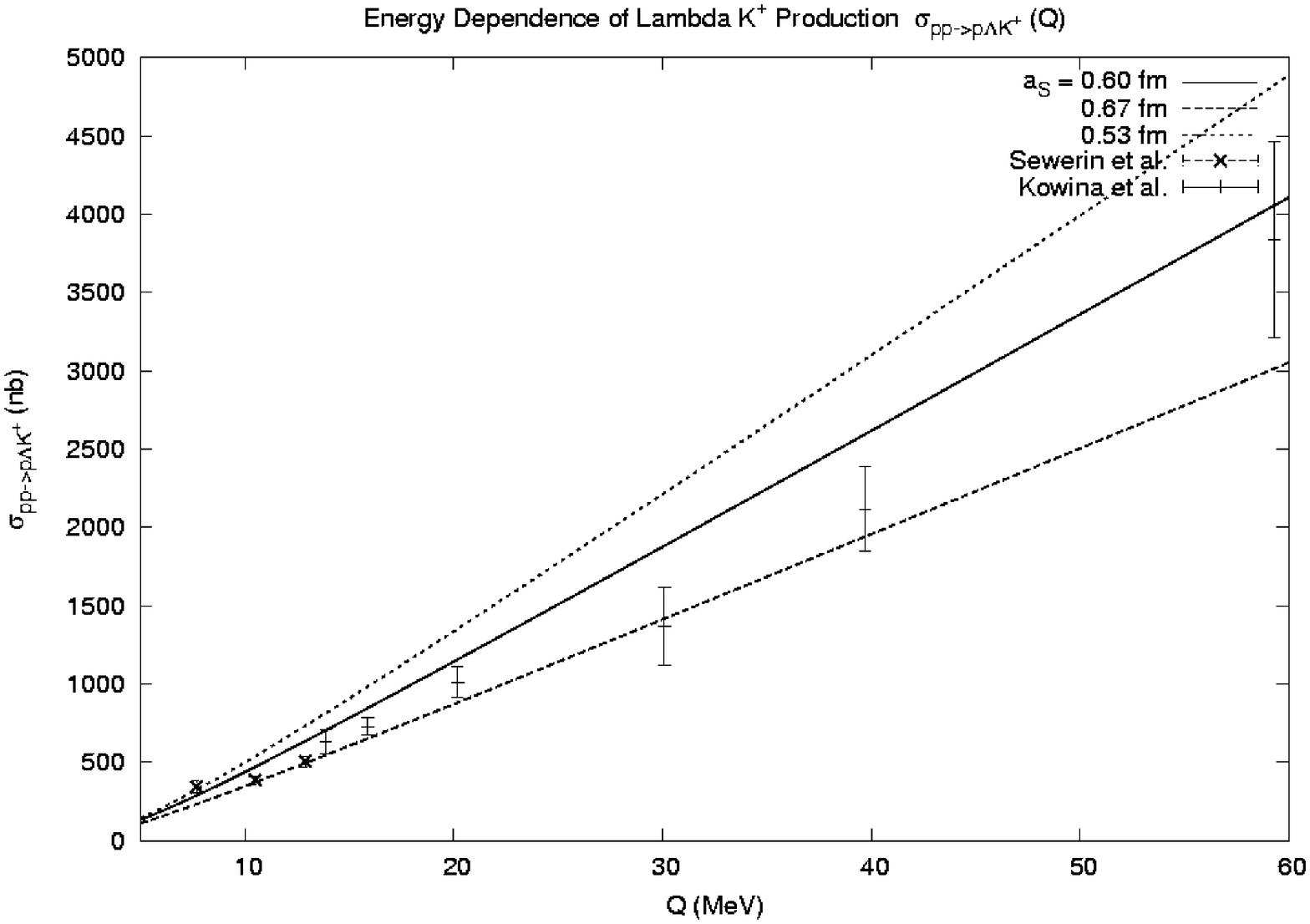}}} 
\framebox{\centerline{\psfig{width=10.0cm,file=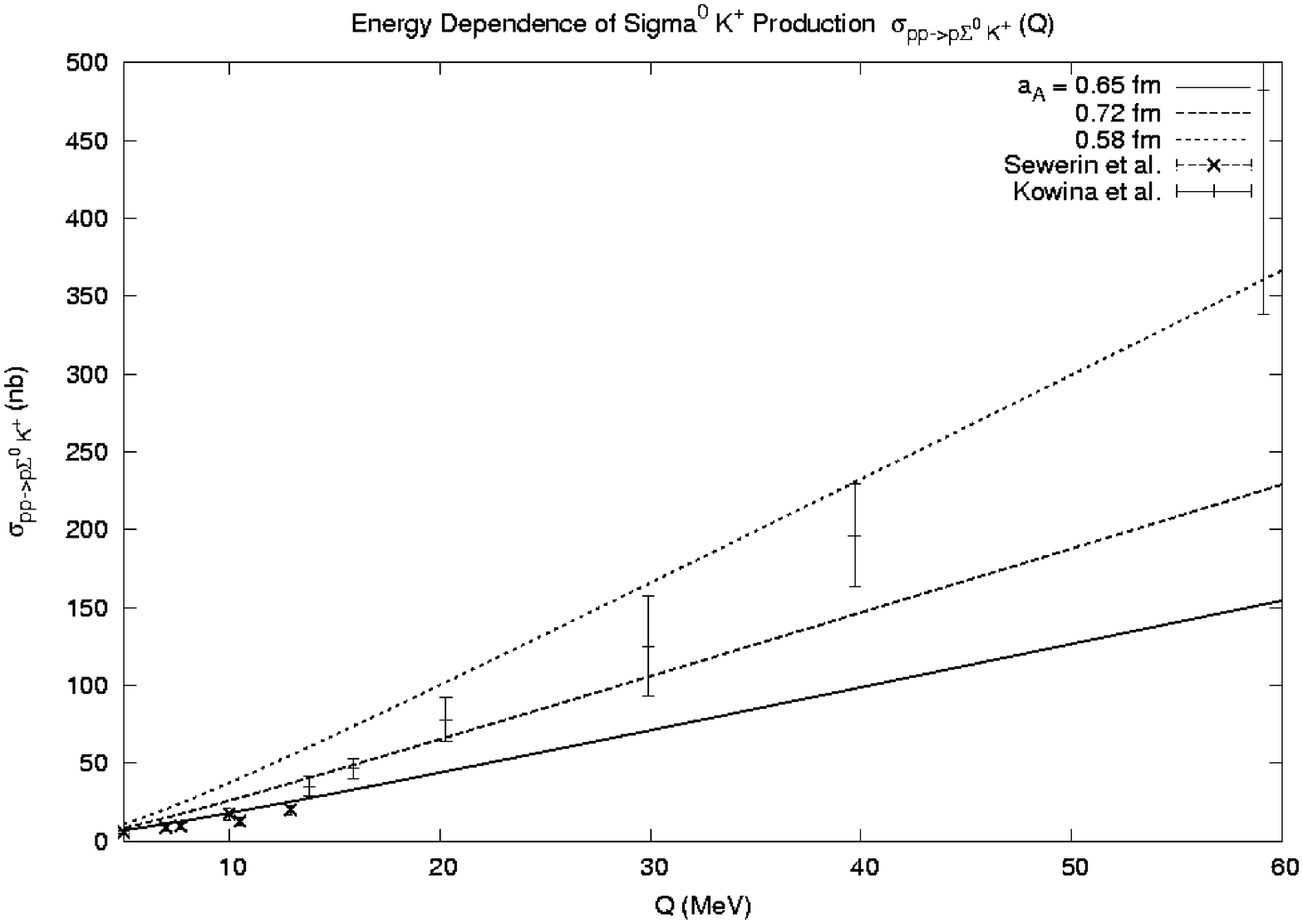}}} 
\vspace*{13pt}
\vspace{-0.5cm}
\fcaption{Q-dependence of the total cross sections $pp \rightarrow p \Lambda K^+ (a)$ and $pp \rightarrow p \Sigma ^0 K^+ (b)$. Compared are results for different size parameters for the scalar diquark in $\Lambda$ production (a) and for the axial diquark in $\Sigma^0$ production (with $P_D$ = 0.1) (b). References see Fig. 1}
\end{figure}

\begin{figure}[htbp] 
\vspace*{13pt}
\framebox{\centerline{\psfig{width=10.0cm,file=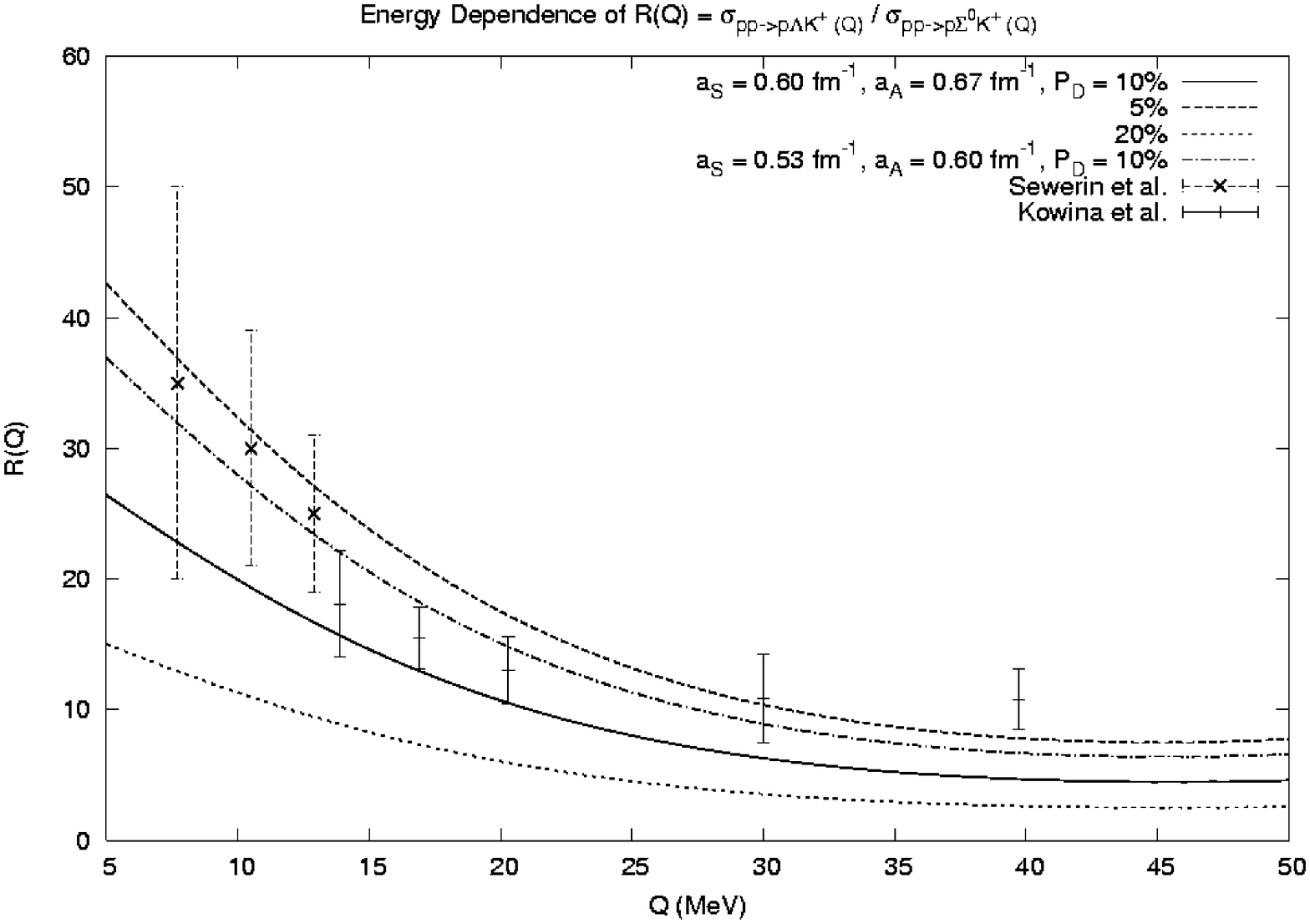}}} 
\vspace*{13pt}
\vspace{-0.5cm}
\fcaption{$R _{\Lambda / \Sigma ^0}(Q)$ for different parameters of the scalar and axial diquark (as specified in the figure and references as in Fig. 1)}
\end{figure}

\end{document}